# AN EXACT RESULT FOR THE PARTITION FUNCTION OF TWO-DIMENSIONAL NEAREST NEIGHBOUR ISING MODELS IN NON-ZERO MAGNETIC FIELD

G.Nandhini and M.V.Sangaranarayanan*

Department of Chemistry

Indian Institute of Technology – Madras, Chennai – 600036 India

**Abstract**

The partition function of two-dimensional nearest neighbour Ising models in a non-zero magnetic field is derived employing a matrix formulation. The deduced partition function values are in agreement with those arising from Onsager's exact solution when $H=0$. The variation of the magnetization with the magnetic field is also provided.

**Introduction**

The analysis of Ising models constitutes a central theme in statistical mechanics and is applied in diverse contexts: the study of critical behaviour of ferromagnets [1]; base pair sequences in DNA [2]. While the analysis of the one-dimensional nearest neighbour Ising models is pedagogical, the study of the two-dimensional Ising models is complex even for vanishing magnetic field [3].An explicit expression for critical temperatures was provided by Lee and Yang [4]. However, subsequent attempts of analyzing the two-dimensional Ising models when the magnetic field is non-zero has remained elusive till now, although methodologies exist viz Bragg-Williams approximation [5], Bethe *ansatz* [6], series expansions [7], renormalization group [8], scaling hypothesis [9]etc. In view of its equivalence with binary alloys and lattice gas description of fluids [10], the results derived in the context of Ising models are also valid for problems in solid state and condensed matter physics. In contrast to the one-dimensional case, the analysis of two-dimensional Ising models employing the transfer matrix method is considered almost impossible since the total number of configurations for a square lattice of N sites is $2^N$ and even for N=16, this number equals 65,536. The two-dimensional Ising model[11] finds applicability in the study of order-disorder transitions [12], electrochemical interfaces [13], phase separation in self-assembled monolayer films [14], protein folding [15], free energies of surface steps [16] etc.

We extend the analysis pertaining to the one-dimensional Ising model and exploit the properties of matrices to derive the canonical partition function Q as

$$Q = \left[ s_0 + \frac{1}{2}\sqrt{s_1 + s_2 + s_3} + \frac{1}{2}\sqrt{2s_1 - s_2 - s_3 + s_4} \right]^N \quad (1)$$

where

$$s_0 = \frac{1}{2}\cosh\left(\frac{2J+H}{kT}\right)$$

$$s_1 = \cosh^2\left(\frac{2J+H}{kT}\right)$$

$$s_2 = \frac{2}{3}\left(\sinh\left(\frac{2J+2H}{kT}\right) + 3\sinh\left(\frac{2J}{kT}\right)\right) \left[ \frac{\sinh\left(\frac{2J}{kT}\right)}{\left(4s_0^2 + \sinh^2\left(\frac{H}{kT}\right)\right) * \left(1 + \sqrt{1 - \left(\frac{4}{3}\right)^3 \frac{\left(\sinh\left(\frac{2J}{kT}\right) + s_0 \sinh\left(\frac{H}{kT}\right)\right)^3}{\sinh\left(\frac{2J}{kT}\right)\left(4s_0^2 + \sinh^2\left(\frac{H}{kT}\right)\right)^2}}\right)} \right]^{\frac{1}{3}}$$

$$s_3 = 2\left\{\sinh^2\left(\frac{2J}{kT}\right)\left(4s_0^2 + \sinh^2\left(\frac{H}{kT}\right)\right)*\left(1 + \sqrt{1 - \left(\frac{4}{3}\right)^3 \frac{\left(\sinh\left(\frac{2J}{kT}\right) + s_0 \sinh\left(\frac{H}{kT}\right)\right)^3}{\sinh\left(\frac{2J}{kT}\right)\left(4s_0^2 + \sinh^2\left(\frac{H}{kT}\right)\right)^2}}\right)\right\}^{\frac{1}{3}}$$

$$s_4 = \frac{4\left[2s_0(2s_0^2 - 1) + \cosh\left(\frac{2J - H}{kT}\right)\right]}{\sqrt{s_1 + s_2 + s_3}}$$

Results and discussion

Partition function

Eqn (1) represents the partition function of the two-dimensional nearest neighbour Ising model in a non-zero external magnetic field for a square lattice of N sites. The values predicted by the above eqn when H = 0 are compared with the Onsager's exact solution in Fig (1). As can be seen, the agreement is satisfactory.

For deducing the partition functions pertaining to $H \neq 0$, we have carried out the complete enumeration of $2^{16}$ configurations pertaining to a square lattice of N sites and counted each energy term in the Hamiltonian. The partition function values arising from this enumeration technique are compared with the values predicted by eqn (1) in Fig 2. There exists complete agreement between the two estimates, suggesting that eqn (1) is indeed the partition function for two-dimensional nearest neighbour Ising model.

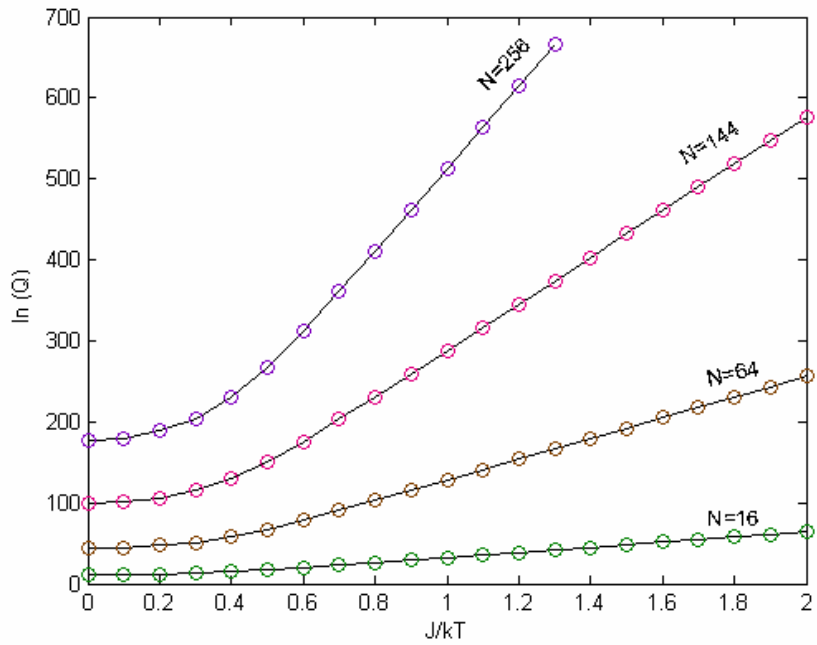

Fig.1: The variation of the partition function with the nearest neighbour interaction energies for various values of N. Lines denote the estimates arising from the Onsager's exact solution while the circles denote the values arising from eqn (1).

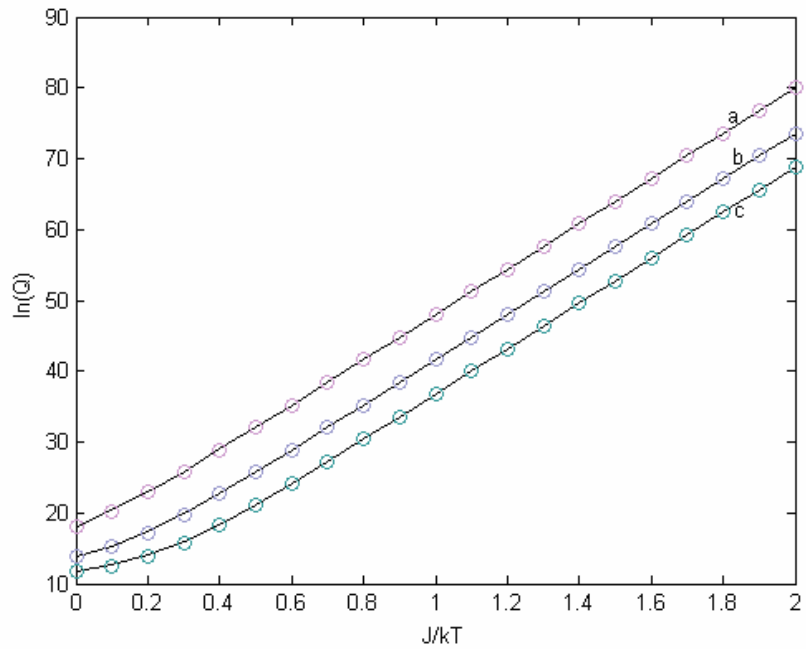

Fig.2 : The partition functions calculated from the equation (1) are denoted by the circles. The values arising from enumeration of $2^N$ configurations are denoted by the lines for N=16.
a. H=1 kT; b. H=0.6 kT; c. H=0.3 kT.

Spontaneous magnetization

The precise definition of the spontaneous magnetization of two-dimensional Ising models is somewhat subtle. Hence we exploit the well-known result of Yang and Lee [4] for critical temperatures and define the magnetization such that at T/Tc =1, the magnetization becomes zero when the derivative of $\left(\dfrac{d \ln Q}{dH}\right)_{H=0}$ from eqn (1) is estimated. Employing this prescription, we depict the behaviour of magnetization as a function of J/kT for different values of H/kT in Fig. 3. Further analysis of the spontaneous magnetization is in progress.

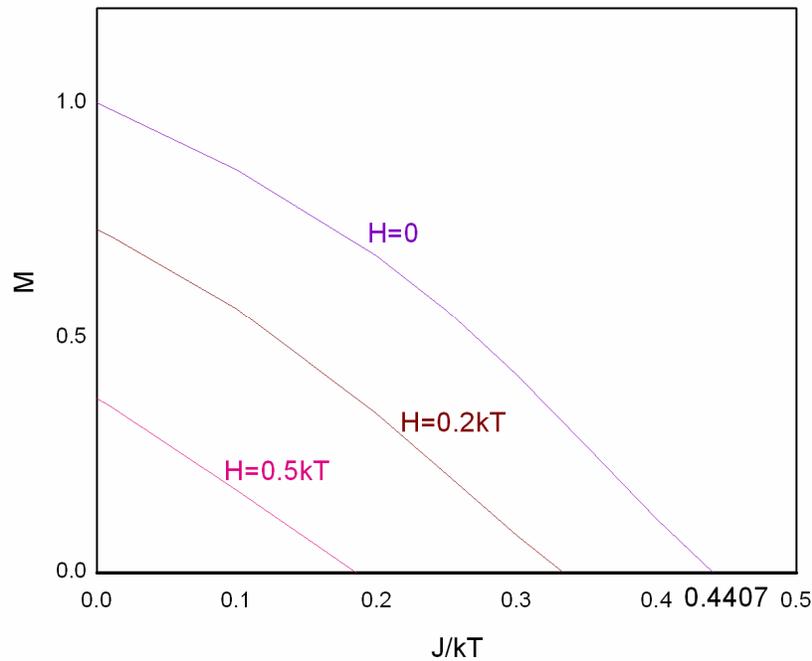

Fig.3 : The dependence of the magnetization on the nearest neighbour interaction energies for various values of the external magnetic field.

Summary

The exact partition function of the two-dimensional nearest neighbour Ising model in a non-zero magnetic field is derived and compared with the computed values. The non-spontaneous magnetization is analyzed for various values of magnetic field and nearest neighbour interaction energies.